\documentclass[twoside,onecolumn,12pt]{article}
\usepackage{extsizes}
\usepackage[super,sort&compress,comma]{natbib} 
\usepackage[version=4]{mhchem}
\usepackage{amssymb}
\usepackage[left=1.2cm, right=1.2cm, top=1.5cm, bottom=2.0cm]{geometry}
\usepackage{balance}
\usepackage{times,mathptmx}
\usepackage{sectsty}
\usepackage{graphicx} 
\usepackage{lastpage}
\usepackage[format=plain,justification=justified,singlelinecheck=false,font={stretch=1.125,small,sf},labelfont=bf,labelsep=space]{caption}
\usepackage{float}
\usepackage{fancyhdr}
\usepackage{fnpos}
\usepackage[english]{babel}
\usepackage{array}
\usepackage{droidsans}
\usepackage{charter}
\usepackage[T1]{fontenc}
\usepackage[usenames,dvipsnames]{xcolor}
\usepackage{setspace}
\usepackage[compact]{titlesec}
\usepackage[normalem]{ulem}
\usepackage{url}
\usepackage[mathscr]{euscript}
\usepackage[running]{lineno}
\usepackage{caption}
\usepackage{setspace}
\usepackage[normalem]{ulem}

\definecolor{cream}{RGB}{222,217,201}
\begin{document}

\pagestyle{fancy}
\thispagestyle{plain}
\fancypagestyle{plain}{\renewcommand{\headrulewidth}{0pt}}

%%%PAGE SETUP - Please do not change any commands within this section%%%
\makeFNbottom
\makeatletter
\renewcommand\LARGE{\@setfontsize\LARGE{15pt}{17}}
\renewcommand\Large{\@setfontsize\Large{12pt}{14}}
\renewcommand\large{\@setfontsize\large{10pt}{12}}
\renewcommand\footnotesize{\@setfontsize\footnotesize{7pt}{10}}
\makeatother

\renewcommand{\thefootnote}{\fnsymbol{footnote}}
\renewcommand\footnoterule{\vspace*{1pt}% 
\color{cream}\hrule width 3.5in height 0.4pt \color{black}\vspace*{5pt}} 
\setcounter{secnumdepth}{5}

\makeatletter 
\renewcommand\@biblabel[1]{#1} 
\renewcommand\@makefntext[1]% 
{\noindent\makebox[0pt][r]{\@thefnmark\,}#1}
\makeatother 
\renewcommand{\figurename}{\small{Fig.}~}
\sectionfont{\sffamily\Large}
\subsectionfont{\normalsize}
\subsubsectionfont{\bf}
\setstretch{1.125} %In particular, please do not alter this line.
\setlength{\skip\footins}{0.8cm}
\setlength{\footnotesep}{0.25cm}
\setlength{\jot}{10pt}
\titlespacing*{\section}{0pt}{4pt}{4pt}
\titlespacing*{\subsection}{0pt}{15pt}{1pt}
%%%END OF PAGE SETUP%%%

%%%FOOTER%%%
\fancyfoot{}
\fancyfoot[RO]{\footnotesize{\sffamily{\thepage}}}
\fancyfoot[LE]{\footnotesize{\sffamily{\thepage}}}
\fancyhead{}
\renewcommand{\headrulewidth}{0pt} 
\renewcommand{\footrulewidth}{0pt}
\setlength{\arrayrulewidth}{1pt}
\setlength{\columnsep}{6.5mm}
\setlength\bibsep{1pt}
%%%END OF FOOTER%%%

%%%FIGURE SETUP - please do not change any commands within this section%%%
\makeatletter 
\newlength{\figrulesep} 
\setlength{\figrulesep}{0.5\textfloatsep} 

\newcommand{\topfigrule}{\vspace*{-1pt}% 
\noindent{\color{cream}\rule[-\figrulesep]{\columnwidth}{1.5pt}} }

\newcommand{\botfigrule}{\vspace*{-2pt}% 
\noindent{\color{cream}\rule[\figrulesep]{\columnwidth}{1.5pt}} }

\newcommand{\dblfigrule}{\vspace*{-1pt}% 
\noindent{\color{cream}\rule[-\figrulesep]{\textwidth}{1.5pt}} }

\makeatother
%%%END OF FIGURE SETUP%%%

%%%TITLE, AUTHORS AND ABSTRACT%%%
\twocolumn[
\begin{@twocolumnfalse}
	\vspace{3cm}
	\sffamily
	\noindent\LARGE{\textbf{Using Multiscale Molecular Dynamics Simulations to Obtain Insights into Pore Forming Toxin Mechanisms\\}} \\%Article title goes here instead of the text "This is the title"
	
	\noindent\LARGE{Rajat Desikan\textit{$^{1}$}$^{\dag}$, Amit Behera\textit{$^{1}$}$^{\ddag}$, Prabal K. Maiti\textit{$^{2}$} and K. Ganapathy Ayappa\textit{$^{1,3}$}$^{\ast}$} \\ \\ \\ \Large{\textit{$^{1}$Department of Chemical Engineering, Indian Institute of Science, Bengaluru, India, 560012 \\ \\ $^{2}$Centre for Condensed Matter Theory, Department of Physics, Indian Institute of Science, Bengaluru, India, 560012 \\ \\ $^{3}$Centre for Biosystems Science and Engineering, Indian Institute of Science, Bengaluru, India, 560012}\\ \\
	\textit{$^{\dag}$Present address: Certara QSP, Certara UK Limited, Sheffield, UK}\\ \\
	\textit{$^{\ddag}$Present address: HP Green R\&D Centre, Hindustan Petroleum Corporation Limited, Bengaluru, India, 560067}\\ \\
	$^{\ast}$Corresponding author, E-mail: ayappa@iisc.ac.in} \\ \\ \\ \\ \\
	
	\noindent \Large \textbf{Document statistics:\newline} 
	\noindent \Large Title: 101 characters including spaces \newline
	\noindent \Large Abstract: 134 words \newline
	\noindent \Large Text: $\sim 8000$ words \newline
	\noindent \Large Figures: 7 \newline
	\noindent \Large Number of tables: 0 \newline
	\noindent \Large References: 147

\end{@twocolumnfalse} \vspace{0.6cm}

]
%%%END OF TITLE, AUTHORS AND ABSTRACT%%%

%%%FONT SETUP - please do not change any commands within this section
\renewcommand*\rmdefault{bch}\normalfont\upshape
\rmfamily
\section*{}
\vspace{-1cm}

\onecolumn
\renewcommand{\figurename}{Figure}

\clearpage
% \newpage

\doublespacing
\raggedbottom
\tableofcontents
% \linenumbers

\clearpage
\newpage
\section*{Abstract}

Pore forming toxins (PFTs) are virulent proteins released by several species, including many strains of bacteria, to attack and kill host cells. In this article, we focus on the utility of molecular dynamics (MD) simulations and the molecular insights gleaned from these techniques on the pore forming pathways of PFTs. In addition to all-atom simulations which are widely used, coarse grained MARTINI models and structure based models have also been used to study PFTs. Here, the emphasis is on methods and techniques involved while setting up, monitoring, and evaluating properties from MD simulations of PFTs in a membrane environment. We draw from several case studies to illustrate how MD simulations have provided molecular insights into protein-protein and protein-lipid interactions, lipid dynamics, conformational transitions and structures of both the oligomeric intermediates and assembled pore structures. \newline \newline 

\noindent {\Large \textbf{Keywords:} Pore forming toxin; membrane protein; lipid bilayer; molecular dynamics simulation; all-atom; coarse-grained; MARTINI; structure-based model}
 
\clearpage
\newpage

\section{Introduction}

Several virulent bacterial strains, such as those causing cholera, pneumonia and listeriosis, initiate their infections by releasing proteins which bind to the target cell membrane to form unregulated membrane-inserted pores, thus compromising ionic balance and eventually causing cell lysis{~\cite{Goot-PFTreview2015}}. These proteins, commonly referred to as pore forming toxins (PFTs), undergo a process of membrane bound oligomerization to form a pore complex, the primary vehicle used to transmit bacterial virulence~\cite{Los2013, Goot-PFTreview2015}. PFTs are classified as either $\alpha$ or $\beta$ toxins depending on the secondary structure of their membrane-inserted helices in the pore state. Pore dimensions depend on the oligomeric state of the final pore complex~\cite{ Goot-PFTreview2015,cosentino2016assembling,Desikan_jbsd_2020}. The pores formed by $\alpha$ toxins are smaller in size consisting of oligomers ranging from 2--26 mers, with some of the largest pores formed by the cytolysin A (ClyA) family of toxins{~\cite{XaxAB,groll2018structurea}}. In contrast, $\beta$ toxins form both smaller and larger pores, with the cholesterol dependent cytolysins (CDCs) forming the largest known PFT pores consisting of 30--50 mers~\cite{Tweten2001}. Upon encountering the membrane, water soluble PFT monomers typically bind using membrane binding motifs and/or membrane receptors, eventually undergoing a conformational change and oligomerization to form the membrane-inserted pore~\cite{Goot-PFTreview2015,rajat-smog}. Understanding the underlying molecular mechanisms of PFT mediated virulence will aid in developing effective treatment strategies to disrupt or compromise pore formation pathways{~\cite{target-PFTs,Escajadillo2018}}, especially in the face of rising antibiotic-resistant bacterial strains{~\cite{antibiotic-resistance,Opal2016}}. 

\begin{figure}[!ht]
\centerline{\includegraphics[scale=0.9]{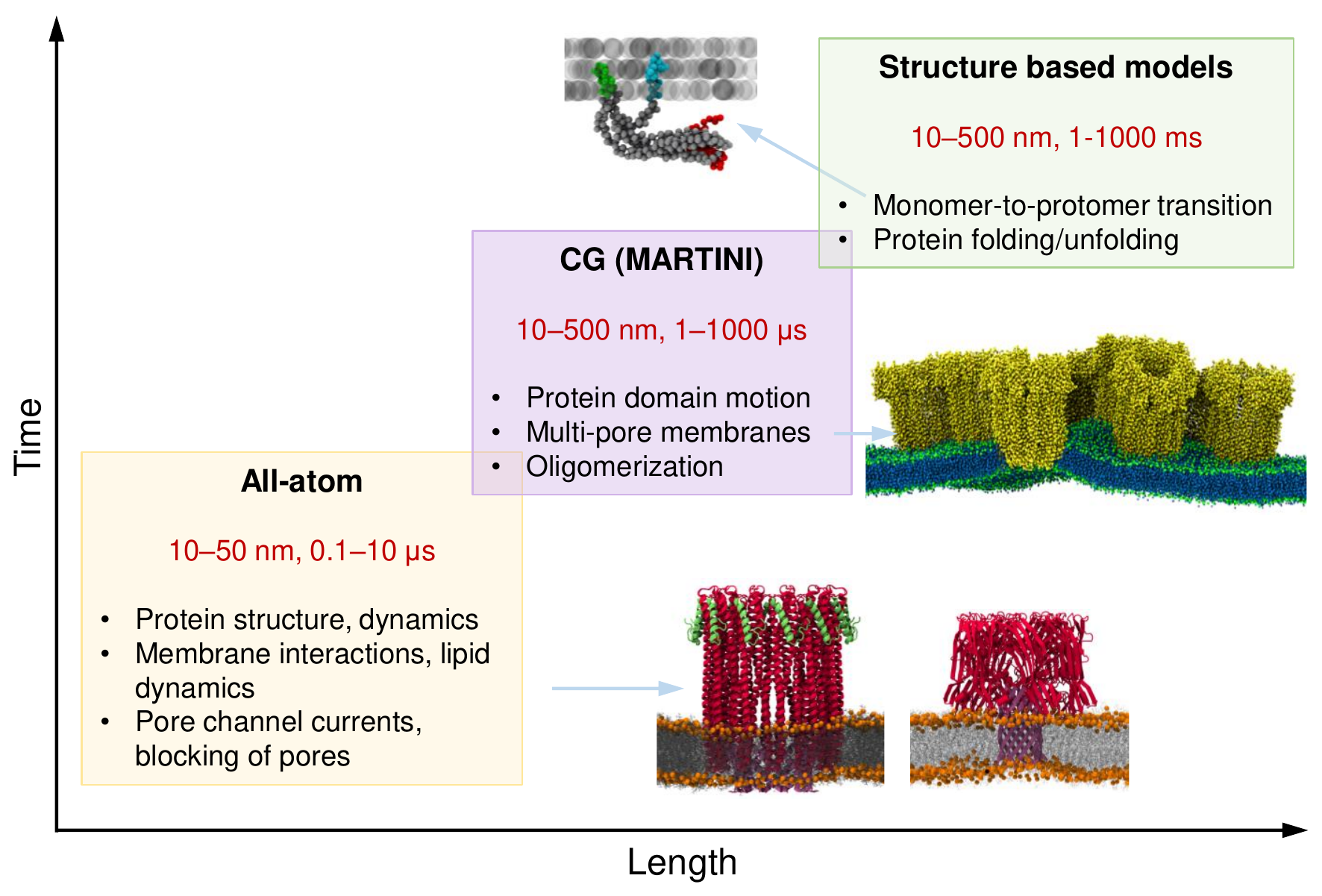}}
 \caption{Illustration of length and time scales typically probed in simulations of pore forming proteins, depicting all-atom, coarse grained (CG) MARTINI and structure based models.}
\label{Fig1}
\end{figure}

In this article, we focus on the utility of molecular dynamics (MD) simulations and the molecular insights obtained from these techniques on the pore forming pathways of several PFTs (Fig.~\ref{Fig1}). A challenge associated with carrying out MD simulations is the large number of atoms associated with the PFT pore complex, as well as their ability to capture slow conformational changes that are an intrinsic part of membrane binding and oligomerization. Unlike most transmembrane proteins, the presence of large extracellular domains of the pore complexes is a characteristic feature of PFTs adding to the computational overhead. (See Ref.~\cite{Desikan_jbsd_2020} for an illustration of 13 $\alpha$ and $\beta$ pore crystal structures with annotated transmembrane domains.) For example, in the case of the ClyA family of PFTs, less than a quarter of the residues are membrane-inserted. Thus, all-atom MD simulations have been sparingly used to understand PFTs. A few notable mentions are simulations studying the dodecameric $\alpha$-PFT, ClyA{~\cite{rajat-cter,rajat-smog,rajat-langmuir,Desikan_pnas_2020,kga-pnas,KGA-nanoscale,Desikan_jbsd_2020,Desikan_SM2020,Desikan2017}}, hexameric assemblies of the membrane attack complex/perforin (MACPF)~\cite{ni2018}, structure~\cite{AHLschulten,Desikan_jbsd_2020,Desikan2017}, lipid dynamics{~\cite{Desikan_SM2020}}, and water/solute transport~\cite{AHLschulten,wong2016,AHL-protein_translocate,AHL-ssdna_translocate,Bonome2019} in the heptameric $\beta$-PFT $\alpha$-hemolysin (AHL) pore, cholesterol sensing motifs for the CDCs streptolysin O (SLO)~\cite{feil2014} and listeriolysin O (LLO)~\cite{ilapnas,ramesh2020}, cry toxin (Crt4Aa) pre-pores~\cite{tavee2010}, peptide sensing in the $\beta$-PFT aerolysin pores~\cite{Degiacomi2013,Cirauqui2017,cao2019}, and conformational changes in Tc toxins~\cite{gatso2018,meusch2014}.

In addition to the all-atom simulations, coarse grained MARTINI models~\cite{Desikan2017,Vogele2019,Desikan_pnas_2020} and structure based models (SBMs)~\cite{rajat-smog,gatso2018} have also been used to study PFTs (Fig.~\ref{Fig1}). Coarse grained methods result in a decrease in the number of degrees of freedom, thus allowing larger system sizes and sampling times compared to all-atom simulations. Each of these methods have their advantages and limitations, and we discuss these with specific examples in this article. All-atom simulations yield detailed insights into molecular interactions, such as, specific cholesterol binding sites{~\cite{kga-pnas,ilapnas}}, stabilizing salt bridges and hydrogen bond networks~\cite{Desikan_jbsd_2020}, and lipid dynamics around pores~\cite{Desikan_SM2020}. The accuracy of these predictions rests with the choice of the inter-atomic potentials used to describe the interactions and forces in the MD simulation (discussed in Section 2.2). In the MARTINI model, protein residues, lipids and water are treated as coarse grained beads suitably modified to describe polar and non-polar interactions. The advantage of the MARTINI models is the ability to simulate large systems sizes which require several tens-to-hundreds of microseconds of sampling time (Fig.~\ref{Fig1}). Both all-atom and MARTINI models are limited in their ability to explore protein conformational transitions that occur during the pore formation pathways of PFTs. SBMs, initially developed to study protein folding~\cite{Clementi2000,SBM-bookchapter}, have been used to capture conformational changes associated with the conversion of a single ClyA monomer to a membrane-inserted protomer~\cite{rajat-smog}.

Here, we focus on the methodology used in our laboratory to study PFTs using various MD methods which include all-atom{~\cite{rajat-cter,rajat-langmuir,Desikan_pnas_2020,kga-pnas,KGA-nanoscale,Desikan_jbsd_2020,Desikan_SM2020,ilapnas,ramesh2020}}, coarse grained MARTINI{~\cite{Desikan2017,Desikan_pnas_2020}}, and SBM simulations{~\cite{rajat-smog}}. In addition to describing the details of these different simulation methods, we also briefly illustrate key properties and molecular insights gained from each method, discuss their advantages and limitations, and end with a perspective.\newline

\section{All-atom MD simulations of PFTs}

In this section, we discuss several practical aspects of setting up atomistic membrane-PFT systems and the specific insights gained from such studies. In particular, we will focus on the $\alpha$ toxin ClyA, as well as the $\beta$ toxins LLO and AHL. For the basics of molecular simulation techniques and analysis, we refer the reader to excellent resources which include both membrane and membrane protein simulations~\cite{Allen2017,Frenkel-Smit,Tamar,Braun2018Best,Grossfield2018Best,Smith2018Simulation,Lemkul2018From,Maginn2018Best,Manna2019_MD_Rev,Enkavi2019_MD_Rev,Membrane-Protein-MD3}.

\subsection{Protein preparation and homology modeling of PFT oligomers and pores}

Prior to embarking on a membrane-PFT simulation, a key pre-requisite is a high resolution structure of the protein. A 12-mer crystal structure of the fully-formed ClyA pore (PDB ID: 2WCD; 3.3 {\AA} resolution), is available in the protein data bank~\cite{mueller}, with unresolved N-terminal and C-terminal residues 1--7 and 293--303, respectively. These amino acids could be important for ClyA's lytic activity as shown by previous mutagenesis studies~\cite{rajat-cter,Atkins,clya-em-wai,ludwig,Oscarsson}. Therefore, these amino acid stretches were homology modelled via the I-TASSER web-server ~\cite{itasser1,itasser2}. Alternatively, the ``Modeller'' or ``Rosetta'' software ~\cite{modeller,Alford2017} could also be employed, as we did for modeling the termini of a single ClyA protomer in Ref.~\cite{rajat-smog}. The modelled residues were then stitched on to the pore crystal structure, and the resulting structure was then assumed to be the complete pore state. All residue protonation states were assumed to correspond to neutral pH. Similarly, a near-complete 7-mer structure of the AHL pore (PDB ID: 7AHL; 1.9 {\AA} resolution), is available in the protein data bank~\cite{song-ahl}, but has residues with missing coordinates for a few atoms, viz, Arg 66 and Lys 70 in chain A, Lys 30 and Lys 240 in chain D, Lys 283 in chain F and Lys 30 in chain G. Since the backbone atom coordinates were available, rather than resorting to I-TASSER or Modeller for full-blown homology modeling, we reconstructed the missing atoms using the ``psfgen'' module implemented in VMD~\cite{namd,vmd}, similar to a previous study that also simulated the AHL pore~\cite{AHLschulten}.

We have also previously simulated the membrane-inserted, assembly-competent, protomer (1-mer) states of ClyA \cite{rajat-langmuir,kga-pnas} and LLO~\cite{ilapnas}. While a crystal structure is available for simulating the fully oligomerized ClyA pore complex, experimental structures of the membrane-inserted protomer, or the partially oligomerized `arcs', which could potentially be intermediates along the pore forming pathway~\cite{rajat-langmuir,Ayush2017}, are currently unavailable. The structure of the ClyA protomer was assumed to be similar to any of the sub-units in the full ClyA pore structure (elaborated in Section 2.3.3)~\cite{rajat-langmuir}. Membrane-inserted protomers generally exist as higher order oligomers – either as fully formed pores or arcs. For example, a single membrane-inserted protomer of ClyA tends to undergo large RMSF fluctuations of the backbone atoms~\cite{rajat-langmuir,ilapnas}. However, for a membrane-inserted dimer~\cite{kga-pnas} or partially oligomerized arcs~\cite{rajat-langmuir}, a significant increase in structural stability, inferred from reduced root mean square fluctuations (RMSF) of the backbone C$_{\alpha}$ atoms, is observed.

In the absence of a crystal structure for the LLO pore, we used the membrane-inserted protomeric state of pneumolysin (PLY, PDB ID: 5LY6; 4.5 {\AA} resolution), which has a 67\% sequence similarity and 87\% structural similarity to LLO, for homology modelling~\cite{vanPee_eLife_PLY}. Using the PLY membrane-inserted state as a template, we employed the SWISS-MODEL web-server~\cite{swissmodel} and obtained a homology modelled LLO protomer structure which showed good global structural alignment with the PLY protomer~\cite{ilapnas}. Ramachandran maps of this homology modelled LLO protomer showed that $\sim$85\% of residues lie in the favoured regions, an improvement over the PLY crystal structure which had $\sim$74\% residues in the favoured regions. Further, upon relaxing the LLO structure in a membrane environment, up to 95\% residues were found to lie in favoured regions, and specific FRET and AFM distances calculated from the structure matched with experiments~\cite{ilapnas}. Therefore, this structure may be a good representation of the LLO protomer, and the above procedure provides a potential computational workflow to obtain protomer PFT structures. 

\subsection{Force fields, simulation settings, and MD protocols}

The accuracy of both protein and lipid force fields (FFs) are important while simulating PFTs. In this regard, all-atom force fields (FFs) from the CHARMM and AMBER families have been extensively used for a wide variety of membrane-protein systems. In our work, to describe the protein and ion interactions, we employed either the AMBER 99SB-ILDN FF{~\cite{amber-ildn}} with $\phi$ corrections{~\cite{thgordon}}, which has been shown to perform well for proteins~\cite{pande-proteinff} and is implemented in the GROMACS MD engine{~\cite{gromacs}} (\url{www.gromacs.org}), or the AMBER12SB FF~\cite{Amber_ff12SB_ff14SB} implemented in the AMBER MD engine~\cite{Amber_MD_2013}. Alternately, we found similar results with the popular CHARMM36 protein+lipid FF~\cite{Charmm36,Charmm36m,Charmm_lipid,Charmm_lipid2} implemented in GROMACS (\url{http://mackerell.umaryland.edu/charmm_ff.shtml#gromacs}, accessed on 16\textsuperscript{th} October 2020) when compared with the AMBER FFs (SI of Ref.~\cite{rajat-langmuir}). We note that it is good simulation practise to compare key properties across FFs to gain confidence in the predictive ability of the atomistic computations. Several FFs exist for lipids, which are important components of membrane-protein systems such as PFTs. One the earliest FFs is the Berger lipid united atom framework~\cite{Berger_lipid1997}, that derives its interaction parameters and charge description from the GROMOS and OPLS FFs. While the original Berger FF did not reproduce the correct electron density or low temperature phase behaviour as seen in experiments, later improvements~\cite{Chiu2009} greatly ameliorated these issues. On the other hand, several fully atomistic lipid FFs such as the AMBER-compatible `Slipid' FF~\cite{slipid}, CHARMM36 lipid FF~\cite{Charmm_lipid,Charmm_lipid2}, and the AMBER family of FFs such as `GAFFlipid'~\cite{Amber_GAFFlipid} and `Lipid14'~\cite{Amber_Lipid14} accurately reproduce several experimentally observed membrane properties such as the area per head group, chain order parameters, bilayer thickness, and lipid diffusivity (description of membrane properties in Refs.~\cite{Lemkul2018From,Desikan_SM2020,Smith2018Simulation}). For our membrane-PFT simulations, we have extensively used the Slipid parameters to model the lipid membrane in GROMACS, in conjunction with the AMBER 99SB-ILDN-$\phi$ FF for protein and ions, and the TIP3P water model{~\cite{tip3p}}. While investigating the propensity of dendrimers to block the ClyA channels~\cite{KGA-nanoscale}, MD simulations were carried out with the AMBER MD software, with the Lipid14 membrane FF used in combination with the AMBER12SB FF for protein and ions, TIP3P model for water, and the generalized AMBER force field (GAFF)~\cite{Amber_GAFF} for the dendrimer. The dendrimer starting structure was built using the dendrimer building toolkit{~\cite{Maingi2012b}}. 

All PFT-membrane systems are charge neutral with either 0.15 M Na$^{+}$Cl$^{-}$/K$^{+}$Cl$^{-}$, which is physiologically relevant, or with only charge balancing counter-ions (0 M salt). The procedure for insertion of PFTs into membranes is described later (Section 2.3.3). For GROMACS simulations, a leapfrog MD integrator~\cite{Allen2017} with an integration time step of 2 fs was used, along with Verlet buffered lists (target energy drift of 0.005 kJ mol$^{-1}$ns$^{-1}$ atom$^{-1}$), and a neighbour list update frequency of once in every 10 steps. All covalent bonds or only h-bonds (depending on the FF), were constrained using the parallelised LINCS~\cite{lincs} or SHAKE~\cite{Shake} algorithms. Electrostatic interactions were numerically computed using the particle mesh Ewald (PME) method~\cite{pme} with a 1.0 nm real space cut-off, and van der Waals interactions were computed with a 1.0 nm cut-off. Except where explicitly mentioned, systems were simulated in the isothermal-isobaric (NPT) ensemble with three-dimensional periodic boundary conditions. Weak temperature coupling was achieved via the stochastic velocity coupling~\cite{vrescale} or the Nos{\'e}-Hoover thermostats~\cite{nose,hoover} (coupling constant of 0.1-0.5 ps). Pressure coupling for the membrane-protein systems was achieved via the isotropic/semi-isotropic Parrinello-Rahman barostat~\cite{parrinello} (coupling constant of 10 ps), with isothermal compressibilities of $4.5\times10^{-5}$ bar\textsuperscript{-1}. To achieve good thermal equilibrium, membrane and/or protein atoms were coupled as one temperature group to the thermostat while the solvent was separately coupled as another temperature group. Visualization of all MD configurations was performed using the VMD software{~\cite{vmd}} (version 1.9.1. and higher).

The initial membrane-protein configurations were relaxed with energy minimization using the steepest descent method, followed by constant monitoring of systemic energies until they plateaued at a minimum. The energy minimized systems were subjected to short NVT and NPT runs with harmonic restraints (force constants of 100-1000 kJ mol$^{-1}$ nm$^{-2}$) on the protein and membrane atoms for achieving thermal equilibration of the solvent and the membrane-protein solvation shells, and were followed by longer unrestrained NPT runs for equilibrating the simulation box fluctuations. A general equilibration recipe is as follows: (i) 0.5–1 ns simulation in the NVT ensemble with harmonic restraints of 1000 kJ mol$^{-1}$ nm$^{-2}$ on all protein and membrane atoms. In GROMACS, the Berendsen thermostat~\cite{Berendsen} can conveniently be employed at this step for quick thermal equilibration~\cite{Lemkul2018From}, though it is known not to produce a true canonical ensemble. Slow heating from 0 K to the reference temperature can also be performed at this step~\cite{KGA-nanoscale}. (ii) 1 ns simulation in the NPT ensemble with harmonic restraints of 1000 kJ mol$^{-1}$ nm$^{-2}$ on all protein and membrane atoms to pack solvent around the membrane-protein atoms, (iii) 1 ns simulation in the NPT ensemble with harmonic restraints of 100-1000 kJ mol$^{-1}$ nm$^{-2}$ on only the protein backbone and membrane phosphorus atoms, and (iv) 1-5 ns unrestrained simulation in the NPT ensemble until the pressure equilibrates. The final configuration from this NPT run can then be utilized for longer production MD simulations. Some of the above equilibration steps could be skipped depending on the system particulars. For instance, ClyA arcs evacuated central lipids within a few ns, necessitating only short restrained equilibration runs lest the phenomenon of interest (central lipid dynamics) be complete within the equilibration phase itself~\cite{rajat-langmuir}.
Confirmation with experimentally available data on the pore structures from atomic force microscopy data in a membrane environment can also be used to validate the simulation methodology~\cite{ilapnas,ramesh2020}.

\subsection{Case studies}
The utility of all-atom simulations are illustrated with specific case studies performed in our laboratory.

\subsubsection{Assessing the structural stability of a PFT monomer from thermal unfolding MD}

Although located in the solvent exposed interface distal from the membrane interface, the C-terminal residues 293-303 of ClyA are essential for its lytic activity~\cite{rajat-cter}. While the deletion of these residues in the ClyA monomer did not alter its ability to bind to target membranes or oligomerize upon addition of detergent, truncated ClyA poorly lysed erythrocytes, and displayed higher susceptibility to proteolysis and thermal unfolding. Since the molecular mechanism of altered activity in the deletion mutant was not immediately forthcoming from experiments, we took cues from thermal unfolding experiments and performed fully atomistic equilibrium and thermal unfolding simulations of wildtype (WT) ClyA monomer and the mutant~\cite{rajat-cter}. 

\begin{figure}[!ht]
\centerline{\includegraphics[scale=1]{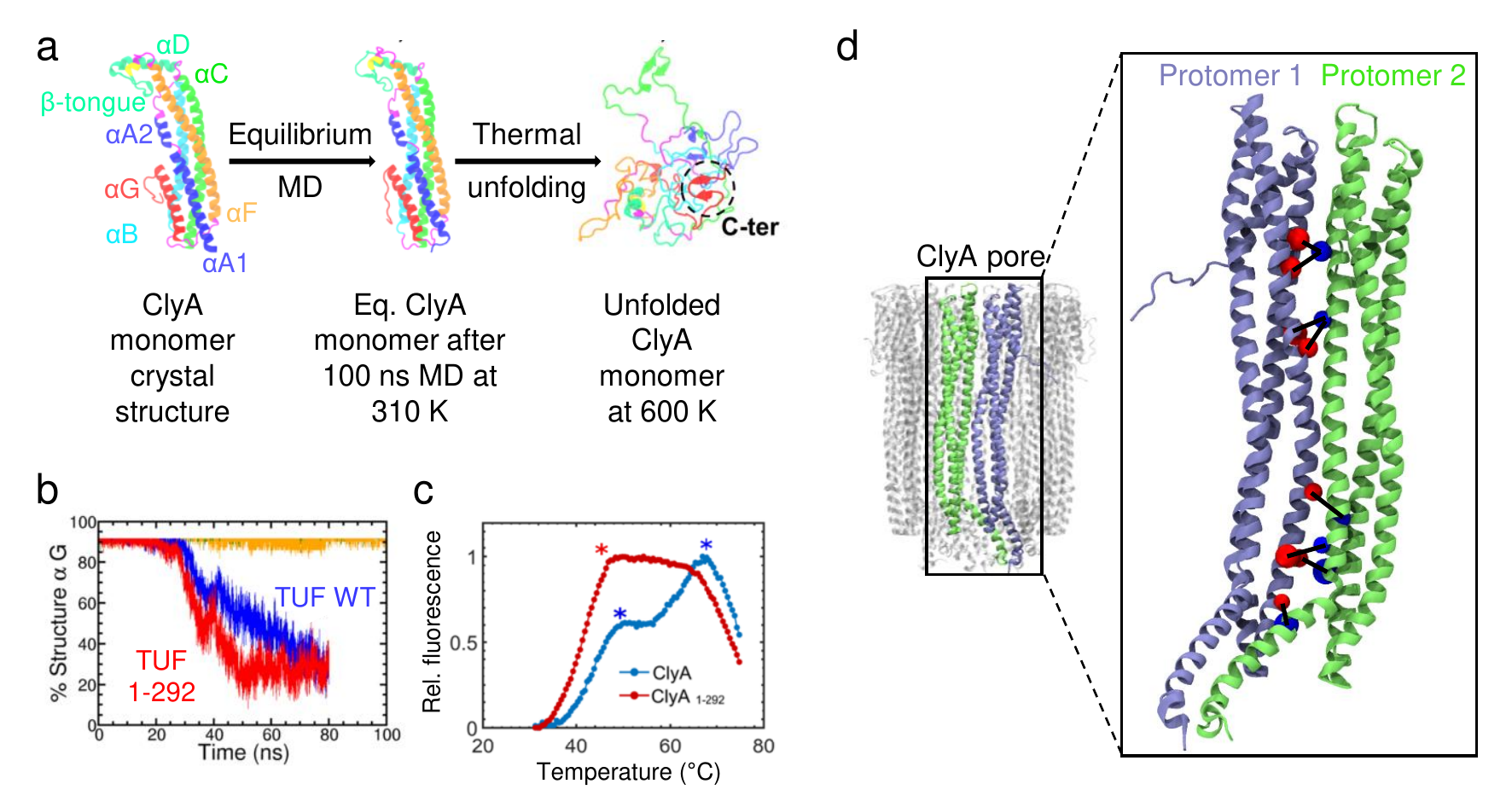}}
 \caption{(\textbf{a-c}) Thermal unfolding MD simulations of the ClyA monomer~\cite{rajat-cter}. (\textbf{a}) Schematic of the thermal unfolding simulations after equilibration at 310 K. (\textbf{b}) Increased loss of helicity in the $\alpha$G helix of truncated ClyA 1--292 compared to wild-type (WT). (\textbf{c}) Thermal denaturation experiments reveal loss of biphasic unfolding for ClyA 1--292 compared to WT. (\textbf{d}) Long lived hydrogen bond and salt bridge networks between two protomers in the ClyA pore complex (details in Ref.~\cite{Desikan_jbsd_2020}). Residues on protomers 1 and 2 depicted as red and blue beads, respectively, and hydrogen bonds/salt bridges as black lines connecting beads. (\textbf{a-c}) Reprinted with permission from {\it Biochemistry} {\bf 55}, 5952–5961 (2016). Copyright (2016) American Chemical Society.}
\label{Fig2}
\end{figure}

The deletion mutant was created by deleting residues 293-298 from the crystal structure of the ClyA monomer (PDB: 1QOY; 2.0 {\AA} resolution), and appropriately capping the C-terminal residue 292 with a negatively charged carboxylate moiety to maintain the zwitterionic state of the protein. The WT structure was built from the crystal structure by modelling missing residues 299-303 similar to the ClyA pore (Section 2.1 and Ref.~\cite{rajat-cter}). Simulations were performed using GROMACS. Unrestrained equilibrium MD simulations of WT and the deletion mutant (three replicates, 100 ns each) were performed in the NPT ensemble at 310 K and 1 bar pressure as described previously (Section 2.2). Global structural analysis from these simulations showed that the deletion mutant was less stable than the WT, especially in the N-terminal helical domains that are proximal to the C-terminus in ClyA's monomer conformation. 
Thermal unfolding MD simulations~\cite{Beck2004,Daggett2002} were performed in the NVT ensemble, and not the NPT ensemble, to ensure that water does not undergo a liquid-to-vapour phase transition with rising temperature (Fig.~\ref{Fig2}a). The final monomer (WT or deletion mutant) configurations, obtained from the aforementioned equilibrium MD simulations, were heated from 310 to 600 K at the rate of $\sim$10 K/ns using the simulated annealing algorithm in GROMACS. Subsequently, a further 50 ns equilibrium MD simulation in the NVT ensemble, at 600 K, was performed to complete the unfolding of the ClyA monomers (Fig.~\ref{Fig2}a). Six independent MD replicates each of WT and the deletion mutant were performed for sufficient sampling of the unfolding pathway, and averages of structural quantities such as the \% helicity of domains (e.g. Fig.~\ref{Fig2}b) were calculated across corresponding replicate trajectories.

%FIGURE2

The deletion mutant exhibited a faster rate of unfolding compared to the WT ClyA monomer, especially in the N-terminal helices that would have been proximal to the C-terminus residues 293-303 (Fig.~\ref{Fig2}). Truncation of these residues could lead to loss of favourable interactions with the N-terminus and therefore result in destabilization of the domain. One of the affected N-terminal helices ($\alpha$A1) eventually forms an integral part of ClyA's transmembrane domain in membrane-inserted states~\cite{mueller,rajat-smog}. The observations from simulations were experimentally validated by thermal unfolding assays~\cite{rajat-cter}, where the deletion mutant completely unfolded at 321 K compared to 341 K for the WT (Fig.~\ref{Fig2}c). Strikingly, WT showed biphasic melting in both MD simulations and experiments, while this behaviour was lost with the deletion mutant in both simulations and experiments (Fig.~\ref{Fig2}b and c). Thus, all-atom MD simulations yield a complementary molecular understanding of the functional role of the C-terminus, indicating that conformational changes required for ClyA pore formation could be compromised by the truncation of the C-terminal residues. These studies suggest that thermal unfolding simulations could serve as a powerful {\it in silico} method for gaining insights into PFT structure, folding, and function.

\subsubsection{Predicting hot-spot residues that stabilize pores via hydrogen bonds and salt bridges}

The sub-units (protomers) of multimeric transmembrane PFT pores are held together by amino acid networks comprising hydrogen bond and salt bridge interactions (discussed in detail in Ref.{~\cite{Desikan_jbsd_2020}}). Identifying key interacting residues, targeting of which could impair pore formation, may potentially enable rational structure-based design of novel therapeutics. While currently available pore crystal structures can be screened \textit{in silico} to identify important inter-protomer stabilizing residues, proteins are flexible and typically exist as a conformational ensemble. Pore crystal structures are a single low-energy static conformation that may not represent the ensemble, and do not contain information regarding protein dynamics. Additionally, most available pore structures are elucidated in detergent media, which can significantly impact protein conformations{~\cite{Chen2019,detinfluenza,detmicelle,detmicelle2}}, and not in lipid membranes which better mimic \textit{in vivo} pore environments. Therefore, instead of resource-intensive experiments such as mutagenesis{~\cite{ahl-CYSscan,AHL-mut1,Mesa-Galloso2017}} and mass spectrometry{~\cite{kallol_gupta_interfacial_lipids}}, all-atom MD simulations of membrane-inserted pore complexes offer an ideal platform to investigate inter-protomer interactions that stabilize PFT pores. 

To analyze these interactions in representative $\alpha$ and $\beta$ toxin pores \textit{in silico}, we created structural ensembles of membrane-inserted ClyA and AHL pores (no lipids in the pore interior) via equilibrium MD simulations, and analysed these ensembles for identifying `hot-spot' inter-protomer interfacial residues engaging in buried hydrogen bonds and salt bridges~\cite{Desikan_jbsd_2020}. We focused on buried hydrogen bonding and salt bridging residues since these interactions are known to facilitate specific protein binding and oligomerization as well as structural stability{~\cite{Desiraju2001,kallol_gupta_interfacial_lipids,sb-buried,sbconcepts,electrostatics1,electrostatics2,sbnetworks}}, specifically in PFTs{~\cite{pnas-wade,mueller}}; the same exercise could also be repeated for hydrophobic interactions. The step-wise simulation protocols are given below.

Using the detergent-stabilized pore crystal structures, we set up membrane-pore systems for MD simulations as specified in Sections 2.1 and 2.2, and simulated each complex for 500 ns at physiologically relevant conditions~\cite{Desikan_jbsd_2020}. By analysing the convergence of global indicators of protein structural stability, we identified parts of the trajectories where both pores were fully equilibrated in membranes. From this converged ensemble, protein dynamics were analysed to obtain occurrence frequencies of hydrogen bonds and salt bridges. Two suitable residues were considered to be hydrogen bonded if the donor-acceptor atomic distance was less than a cut-off threshold of 3 {\AA}, and the hydrogen-donor-acceptor angle was less than a cut-off threshold value of 20$^{\circ}$. Similarly, a salt bridge occurred if any oxygen atom of an acidic amino acid (glutamic acid, aspartic acid) was within a cut-off threshold distance of 4\,{\AA} from any nitrogen of a basic residue (arginine, lysine, histidine). Occurrence frequencies were defined as the fraction of pore conformations in the equilibrium MD ensemble where a particular hydrogen bond or salt bridge was switched on. A small subset of persistent (> 50\% occurrence), buried (relative solvent accessibility < 0.25), inter-protomer, strongly interacting (interaction energies < -4 kcal/mol) hydrogen bonds and salt bridges were identified (see Ref.~\cite{Desikan_jbsd_2020} for more details of the screening procedure). Fig.~\ref{Fig2}d illustrates several long lived salt bridges in the extracellular and membrane-inserted segments for the ClyA pore complex. Previous mutagenesis studies{~\cite{kga-pnas,pnas-wade,AHL-mut1,AHL-mut2,ahl-CYSscan}}, where mutating some of these important residues involved in long lived hydrogen bonds and salt bridges led to diminished pore formation and lytic activity, validated our observations.

\subsubsection{Setting up PFT `arcs' and intermediates in lipid membranes and analysis of central lipid evacuation}

In addition to simulations with the full dodecameric ClyA pore, we have also performed simulations of the single transmembrane protomer (1-mer), and intermediate ClyA `arcs' at various levels of oligomerization (6-mer to 10-mer)~\cite{rajat-langmuir}. In the absence of high-resolution experimentally determined structures of any PFT arc, we constructed the ClyA arc structures from the fully formed pore crystal structure and tested their influence on lipid bilayers to glean molecular insights regarding pore forming mechanisms~\cite{rajat-langmuir}. Similar procedures were also employed to simulate PLY and LLO pore intermediates in membranes to unravel mechanisms of pore opening~\cite{Vogele2019,ramesh2020}. 

\begin{figure}[!ht]
\centerline{\includegraphics[scale=1]{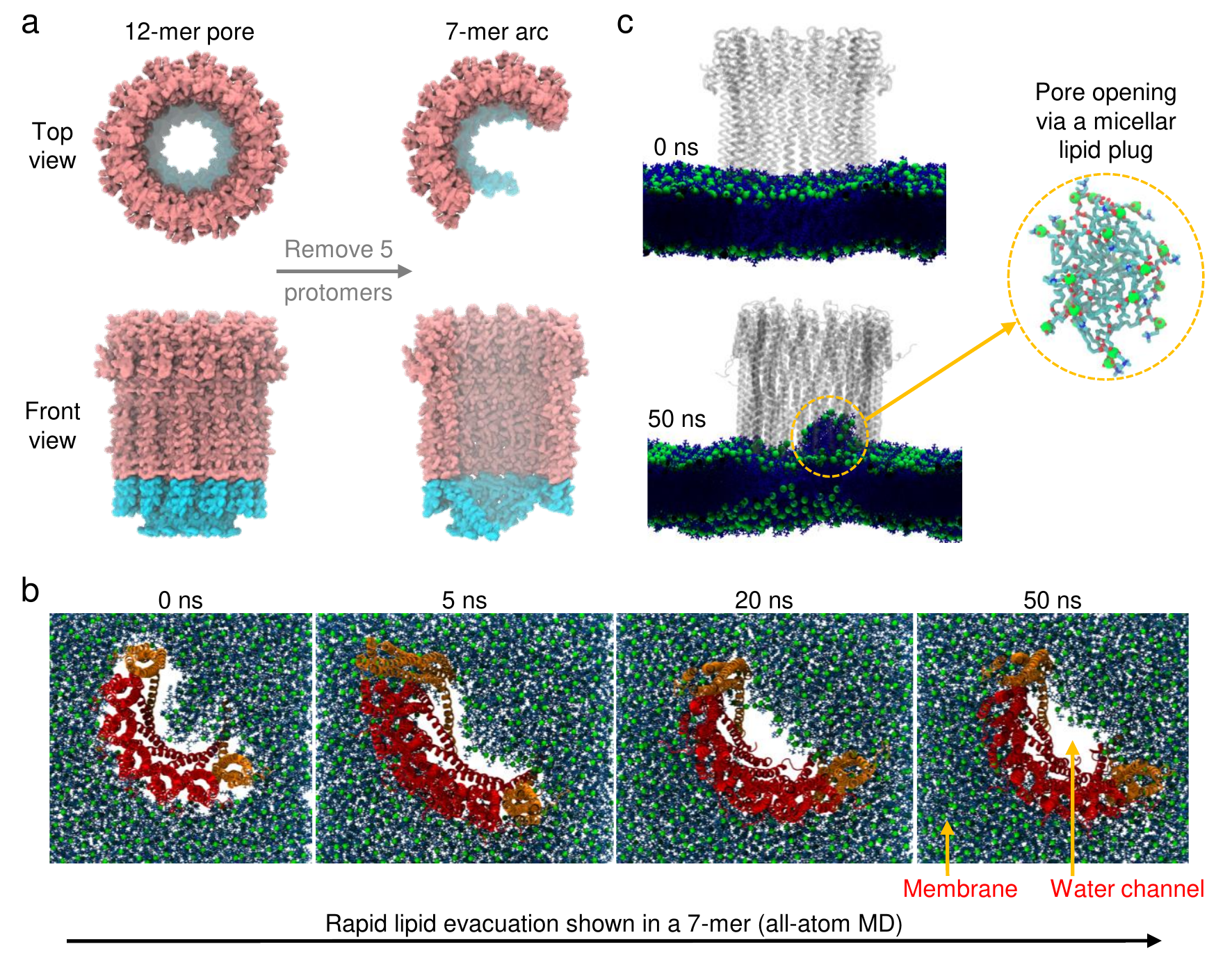}}
 \caption{Molecular dynamics simulations of pore opening with ClyA oligomers/pore~\cite{rajat-langmuir,Desikan_pnas_2020}. (\textbf{a}) Creating initial structures of ClyA arcs from the pore illustrated for the 7-mer. (\textbf{b}) Lipid evacuation from the 7-mer arc interior to create functional pores capable of leakage~\cite{rajat-langmuir}. (\textbf{c}) Micellar lipid plug formation within a fully oligomerized ClyA pore complex~\cite{Desikan_pnas_2020}. (\textbf{a,b}) Reprinted with permission from {\it Langmuir} {\bf 33}, 11496-11510, (2017). Copyright (2017) American Chemical Society. (\textbf{c}) Reproduced from {\it Proc. Natl.Acad. Sci. U.S.A}, {\bf 117}, 5107-5108 (2020).} 
\label{Fig3}
\end{figure}

The structures of the ClyA pore intermediates, from the protomer (1-mer) to the partial n-mer arcs (n=6…10), were obtained from the aforementioned 12-mer pore structure with added termini (Section 2.1), by deleting the requisite number of protomers to create an arc. For example, the 7-mer arc was obtained by deleting 5 adjacent protomers from the 12-mer pore structure (Fig.~\ref{Fig3}a). These n-mer arcs were then manually inserted into fluid-phase DMPC (1,2-dimyristoyl-sn-glycero-3-phosphocholine) and POPC (1-palmitoyl-2-oleoyl-sn-glycero-3-phosphocholine) lipid membranes, by carefully orienting their putative transmembrane iris-like $\alpha$-helical protein domains within the membranes and subsequently deleting lipids that overlapped with the protein. We note that lipids were present in and around ClyA arcs at the start of simulations, and the lipid bilayers were intact and defect-free except for the inserted n-mer complexes. We note that an alternate method for constructing membrane-inserted protein configurations is via the CHARMM‐GUI web-server~\cite{charmmgui1,charmmgui2,charmmgui3}. 

After energy minimization and short restrained equilibration runs (see Section 2.2), these membrane-protein systems were simulated for up to a few hundreds of nanoseconds. Protein structural attributes, lipid dynamics, and membrane rearrangements were continuously tracked.

%FIGURE3

Simulations of the ClyA protomer and the n-mer arcs showed that they were all structurally stable in the membrane over simulation timescales~\cite{rajat-langmuir}, implying that they were reasonable approximations to putative structures of pore intermediates~\cite{Ayush2017,peng2019high}. Strikingly, all arcs, including the protomer, spontaneously displaced the central lipids within the arc interior (shown for the 7-mer in Fig.~\ref{Fig3}b), thus creating protein-lipid-lined transmembrane water channels whose size was commensurate with the extent of PFT oligomerization. The kinetics of lipid evacuation from arc interiors, measured using the continuous survival probability formalism (see Eq.~2), was rapid and was typically completed within tens of nanoseconds.

All-atom simulations of the ClyA membrane-pore complex typically involve about 10$^5$-10$^6$ atoms for the full pore complex. In contrast, larger PFTs such as those from the CDC family, which form oligomers of variable size containing 30-50 protomers, can result in systems involving up to 10 million atoms, thus drastically increasing the computation overhead. In these situations, reduced systems with smaller oligomers can be judiciously constructed to yield insights into the pore forming mechanism. After homology modelling of the LLO membrane-inserted pore state, we studied two tetrameric ensembles, one consisting of four membrane bound D4~\cite{Tweten2001,Vogele2019} sub-units, and the other consisting of four protomers to form a membrane-inserted arc (Fig.~\ref{Fig4}e). To construct the arc, a single membrane-inserted protomer was translated by one unit cell and rotated by 20$^\circ$ with respect to its neighbour to generate the initial LLO tetrameric structure~\cite{koster2014crystal}. Initialization and equilibration procedures were similar to those described in Section 2.2. The D4 sub-units assembled into a linear array on the membrane consistent with electron micrograph data~\cite{koster2014crystal}. Similar to what was observed with ClyA~\cite{rajat-langmuir} and PLY~\cite{Vogele2019} arcs, lipids were rapidly displaced from the arc interior to reorient and form a toroidal interface (Fig.~\ref{Fig4}e). Interestingly, lipids in the toroidal interface were devoid of cholesterol.

In summary, all-atom MD simulations provide strong evidence that partially oligomerized membrane-inserted arc-like states are stable entities for both $\alpha$-PFTs as well as larger CDCs such as LLO and PLY. These simulations also unravelled structural and dynamical molecular features of membrane-inserted pore intermediates which are difficult to investigate experimentally.

\subsubsection{Simulating fully formed pores with intact central lipids to explore pore opening}

We subsequently explored alternate molecular mechanisms of ClyA pore formation and central lipid ejection via MD simulations. Our previous ClyA arc MD simulations were initialized with membrane-inserted arcs, and we had not explored how arcs would insert into intact membranes. A recent study~\cite{Vogele2019} of PLY, with a coarse grained MARTINI FF, showed that with slow insertion of PLY arcs, central lipids could laterally escape from the arc interior, similar to our previous observations with ClyA arcs~\cite{rajat-langmuir}. However, during rapid insertion of PLY arcs, the central lipids in the lumen were trapped, and spontaneously buckled forming a liposomal plug which could then be expelled by osmotic flows. PLY pores/arcs have a large internal diameters of 30 to 50 nm, permitting the liposomal plug-formation pathway. How smaller pores such as ClyA, with internal diameters of $<$ 10 nm~\cite{mueller,Desikan2017}, would expel central lipids was not understood.

Therefore, similar to the arc simulations with ClyA, we inserted fully formed 12-mer ClyA pores into intact DMPC and POPC membranes, and deleted only the lipids that overlapped with the protein~\cite{Desikan_pnas_2020}. This initially left the central lipids in an intact lamellar configuration within the pore lumen (Fig.~\ref{Fig3}c). After energy minimization and short equilibration runs, production MD simulations were performed, and the central lipids monitored. Remarkably, within a few nanoseconds, the central lipids within the pore lumen spontaneously formed stable micellar (not liposomal) lipid plugs, that slowly rose from the membrane plane within the pore channel, and could presumably be expelled later via osmotic flows (Fig.~\ref{Fig3}c). The micelles exhibited solvated phosphatidylcholine head groups and desolvated lipid hydrocarbon tails (Fig.~\ref{Fig3}c), similar to lamellar configurations, and indicative of stability via the hydrophobic effect. Our work suggested that both small and large PFT nanopores may exhibit central lipid ejection as micelles and liposomes, respectively, during pore formation~\cite{Desikan_pnas_2020,Vogele2019}.

\subsubsection{Lipid and cholesterol binding and dynamics in PFTs}

\begin{figure}[!ht]
\centerline{\includegraphics[scale=1]{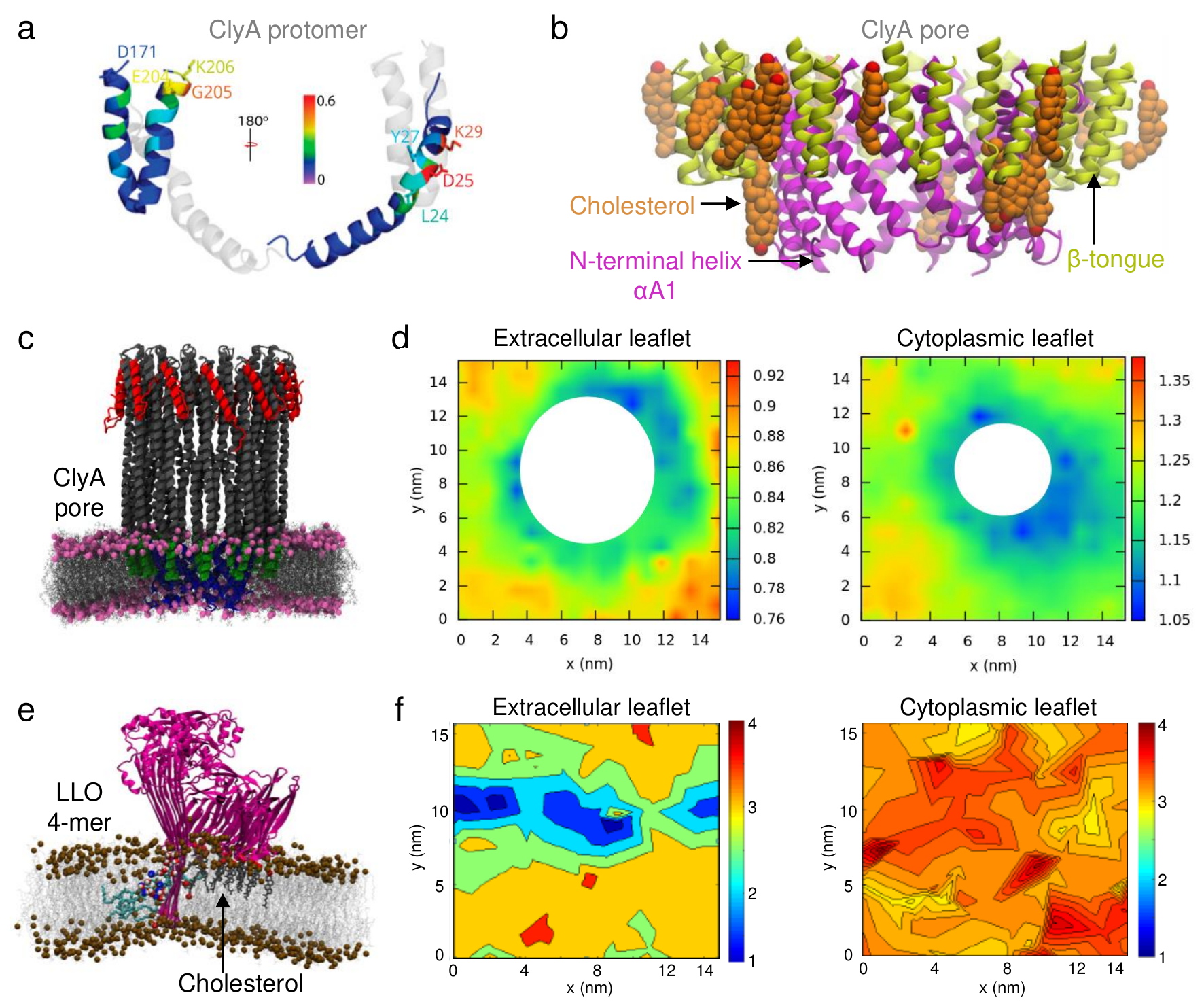}}
 \caption{All-atom MD simulations reveal (\textbf{a}) cholesterol binding sites in a single membrane-inserted ClyA protomer and (\textbf{b}) cholesterol binding pockets between adjacent $\beta$-tongues in the transmembrane helices of the ClyA pore~\cite{kga-pnas}. Simulations are carried out for (a) the entire protomer and (b) the dodecameric pore complex in solvent and membrane, and only the cholesterol binding regions are displayed here for clarity. From MD simulations of a transmembrane ClyA pore (\textbf{c}), lipid displacement, $d_n$ (Eq.~\ref{eq:Mobility}) maps in a DMPC membrane with ClyA (\textbf{d}) illustrate regions of lowered mobility in the vicinity of the pore complex~\cite{Desikan_SM2020} (color bars are in units of nm, $\Delta t$ = 50 ns). Similarly for the LLO tetramer~\cite{ramesh2020} (\textbf{e}), maps of $d_n/\Delta t$ for cholesterol in a DOPC--Cholesterol bilayer (70:30) are shown (\textbf{f}) (color bars are in units of\, \AA/ns, $\Delta t$ = 1 ns), illustrating reduced lipid mobility in the extracellular leaflet due to specific binding with the D4 sub-units as seen in the snapshot (e). (\textbf{a, b}) Reproduced from {\it Proc. Natl. Acad. Sci. U.S.A}, {\bf 115}, E7323-E7330 (2018), (\textbf{c, d}) reproduced from {\it Soft Matter}, {\bf 16}, 4840–4857 (2020).}
\label{Fig4}
\end{figure}

Studies have shown that presence of cholesterol in lipid membranes improves the binding and oligomerization of PFTs. With ClyA, cholesterol is known to enhance activity{~\cite{kga-pnas,Oscarsson,Goot-PFTreview2015}}. To understand the role of cholesterol in ClyA pore formation, all-atom MD simulations of a single protomer, dimer, and the fully oligomerized dodecameric pore complex of ClyA, embedded in DOPC--cholesterol (70:30) membranes, were carried out~\cite{kga-pnas} (Fig.~\ref{Fig4}a and b). For studies with LLO, microsecond-long all-atom MD simulations were performed with a tetrameric membrane bound D4 sub-units as well as membrane-inserted tetrameric arcs~\cite{ramesh2020} (Fig.~\ref{Fig4}e).

Binding is assumed to occur when the distance between the center of masses of cholesterol and the specific residue lies within 0.5 nm~\cite{kga-pnas}. We evaluated the fraction of MD sampling time a cholesterol molecule spends in the vicinity of specific residues. Amino acids K29, Y27 and D25 (Fig.~\ref{Fig4}a), belonging to the cholesterol recognition and consensus (CRAC) motif~\cite{rajat-smog}, were identified as strong cholesterol binding sites~\cite{kga-pnas}. Additional binding sites were also identified between the $beta$-tongues of the assembled pore complex (Fig.~\ref{Fig4}b). Erythrocyte experiments with single mutations of ClyA -- K29A, Y27A, and Y27F -- indicated that Y27 plays a key role in cholesterol recognition~\cite{kga-pnas}. Interestingly Y27 is also part of a long lived hydrogen bond (Fig.~\ref{Fig2}d; Ref.~\cite{Desikan_jbsd_2020}) stabilizing the membrane-inserted N-terminus in the assembled pore complex.

Molecular dynamics simulations of a DMPC bilayer with ClyA and AHL highlight PFT induced changes in lipid dynamics, and the spatial extent of PFT influence on membrane dynamics and lipid structure~\cite{Desikan_SM2020}. Different diffusive regimes were elucidated by the mean square displacements (MSD) calculated using, 
\begin{equation}
	\langle MSD \rangle= \frac{1}{N} \langle \sum_{i=1}^{N} | {\mathbf{r}}_i(t+\Delta t)-\mathbf{r}_i(t)|^2 \rangle
	\label{eq:MSD}
\end{equation}
where, $\mathbf{r}_i = \mathbf{e}_x x_i + \mathbf{e}_y y_i$ and the brackets indicate shifted time averaging. While evaluating diffusion coefficients, one needs to ensure that the particle dynamics have been sufficiently sampled to enter the diffusive regime where MSD $\sim$ t$^{\alpha}$ with the exponent $\alpha = 1$. In the sub-diffusive regime $\alpha < 1$~\cite{Desikan_SM2020,Maginn2018Best}. Sufficiently long MD simulations of at least 1 $\mu s$ must be carried out to reliably extract a diffusion coefficient. Additionally diffusion coefficients are colored by periodic boundary conditions~\cite{vogele2018hydrodynamics}. Despite these caveats, lipid diffusion coefficients extracted from MD simulations compare reasonably well with fluorescence correlation spectroscopy (FCS) measurements~\cite{ilapnas,ramesh2020}. 

The continuous survival probabilities (CSP)~\cite{rajat-langmuir,Desikan_SM2020}, which indicates the duration a lipid molecule resides in a given region is computed using,
\begin{equation}
	C_j(t)= \sum_{i=1}^{N_j}\langle \prod_{t_k=t_0}^{t_0 + t} \delta_i (t_k) \rangle
	\label{eq:CSP}
\end{equation}
where $\delta_i$ is the Kronecker delta, which is either unity if the co-ordinates of the phosphate head group of the $i$\textsuperscript{th} lipid are present in the corresponding region, or zero if absent. The normalized CSP at any time `$t$', $C_j (t)/C_j (t_0)$, yields the fraction of time a molecule is continuously present in a given region~\cite{Desikan_SM2020,rajat-langmuir}. The extent of dynamic perturbation as revealed from the CSPs extend to greater distance ($\sim$ 4 nm) from the pore complex when compared with the structural perturbations which are shorter ranged ($\sim$ 2.5 nm)~\cite{Desikan_SM2020}.

Occupation of cholesterol around the pore can be quantified by using number density maps. The lipid bilayer area is divided into a 2D-grid, with grid lengths of 10 {\AA}. The intersecting points of grid lines is a grid point. The coordinates of the center-of-mass of each cholesterol molecule in the membrane plane was calculated, and the molecule was assigned to the nearest grid point. Grid point counts were then averaged over the simulation trajectory and normalized with the area of the box to obtain the area density. The densities were then interpolated using MATLAB routines using the `scattered Interpolant' function for plotting contour maps. The density maps for cholesterol for both ClyA~\cite{kga-pnas} and LLO~\cite{ramesh2020}, revealed the presence of distinct hot spots for cholesterol binding. 
 
A convenient metric to determine the extent of spatial variation or dynamical heterogeneity in lipid and cholesterol dynamics around the pore complex is the in-plane displacement of a given particle, 
\begin{equation}
	d_{n} = \langle|\mathbf{r}_{n}(t+\Delta t)-\mathbf{r}_{n}(t)| \rangle \\
	\label{eq:Mobility}
\end{equation}
where $\mathbf{r}_n(t) = x_{n}(t){\mathbf e}_{x} + y_{n}(t) {\mathbf e}_{y}$ and 
$\Delta$t is the specific time interval between two configurations, and brackets indicate a time average. The displacements are binned on a two-dimensional grid of 1 {\AA} grid spacing based on the location of the initial co-ordinates of the displacement vectors. As the initial co-ordinates of the displacement vectors were not uniformly distributed, the `ScatteredInterpolant' function in MATLAB was used to interpolate the scattered data to a uniform grid, and the interpolated values on the query points were used for creating contour plots or maps. Fig.~\ref{Fig4}d illustrates the lipid displacement maps for lipids around ClyA, and Fig.~\ref{Fig4}f illustrates the $d_{n}/\Delta t$ maps for cholesterol around LLO tetramers, showing reduced mobility of lipids and cholesterol as well as a distinct variation across leaflets for LLO. This dynamical heterogeneity induced by LLO in DOPC--Cholesterol membranes result in two sub-populations of diffusing lipids as revealed by super resolution measurements~\cite{sarangi2016super}.

The local density and displacement evaluations from MD simulation trajectories established that target cell membranes are not just frameworks for binding of PFTs, but contain molecular determinants that also regulate the activity of the PFTs. Molecular dynamics simulations elucidated the stabilising interaction of cholesterol with ClyA at various stages of the pore forming pathway, namely, the single membrane-inserted protomeric state, membrane-inserted dimers, and the fully assembled dodecameric pore~\cite{kga-pnas}. Similarly, the role of cholesterol in LLO pore forming activity and the induced modulation of lipid and cholesterol density was also established~\cite{ilapnas}.

\subsubsection{Pore blocking and ionic currents in PFT pores}

We have also explored the use of dendrimers as efficient pore blockers to prevent PFT function~\cite{KGA-nanoscale}. Dendrimers are a class of hyper-branched polymers with well-defined size and shape which can be tuned by solution pH{~\cite{Maiti2005,Maiti2009}}, and have shown promising application in drug delivery and therapeutics{~\cite{Maingi2012a,Jain2013}}. All-atom MD simulations were used to compare the pore blockage characteristics between the non-protonated (G5-NH2) polyamido amine (PAMAM) dendrimer and the protonated (G5-NH3$^+$) PAMAM dendrimer~\cite{KGA-nanoscale}. Atomistic MD simulations were carried out using the PMEMD module of the AMBER41 software package. The energy minimized structure was gradually heated up from 0 to 300 K, while imposing a weak harmonic constraint of 20 kcal mol$^{-2}$\,{\AA}$^{-2}$ on the solute atoms. The heated system was simulated for 100 ns in the NPT ensemble under ambient conditions. Protonated (P) and non-protonated (NP) PAMAM were built using the dendrimer building toolkit{~\cite{Maingi2012b}} and energy minimised using the same above mentioned protocol using the Generalized Amber force field (GAFF). The equilibrated dendrimer structure was placed near the extracellular end of ClyA pore opening through VMD and followed by solvation \& addition of ions for maintaining a 150 mM salt concentration. The entire system was then simulated for 200 ns.

\begin{figure}[!ht]
\centerline{\includegraphics[scale=1]{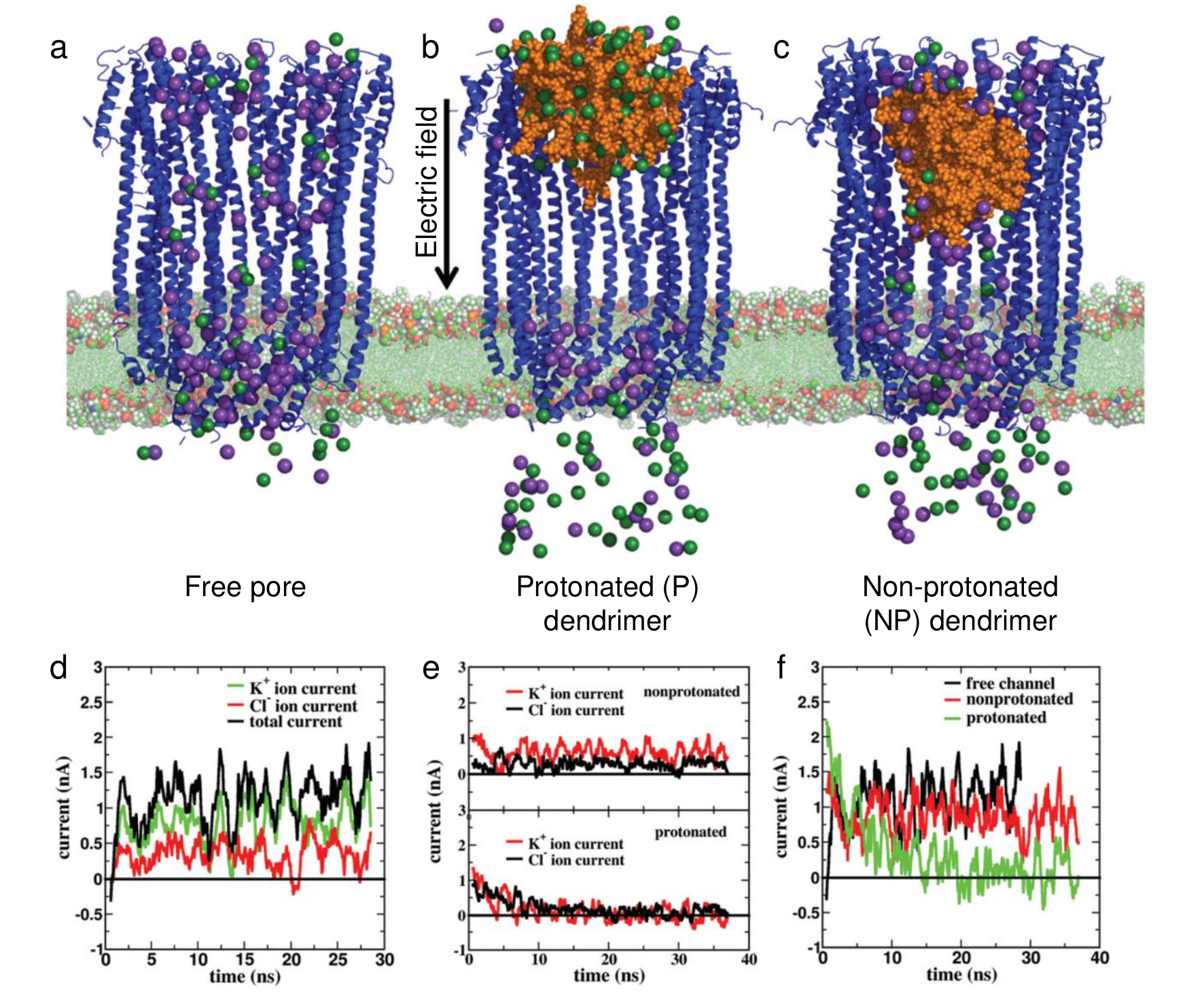}}
 \caption{All-atom MD simulations illustrating ClyA pore blocking with protonated and non-protonated PAMAM dendrimers~\cite{KGA-nanoscale}. Snapshots illustrate (\textbf{a}) ions in a free pore, and effective pore blocking with (\textbf{b}) the protonated dendrimer and (\textbf{c}) the non-protonated dendrimer. Ionic currents in (\textbf{d}) free pores, (\textbf{e}) pores with dendrimers, and (\textbf{f}) the total ionic current with dendrimers are also shown. Cl$^{-}$ -- green, K$^{+}$ -- purple. Reproduced from {\it Nanoscale}, {\bf 8}, 13045-13058 (2016).}
\label{Fig5}
\end{figure}

Ionic current along the ClyA protein channel were induced by imposing a uniform electric field of 0.02 V nm$^{-1}$ directed toward the cytosolic end of the pore. During the course of 35 ns MD simulations with the electric field, the bilayer was restrained to its initial configuration with a harmonic potential. The instantaneous ionic current, $I(t)$, was measured using the following equation,
\begin{equation}
	I(t)= \frac{1}{\Delta tL_z} \sum_{i=1}^{N}q_i[z_i(t+\Delta t)-z_i(t)],
	\label{eq:ion current}
\end{equation}
where $z_i$ and $q_i$ are the $z$ coordinate and atomic charge of the $i^{th}$ atom, respectively, and $N$ represents the total number of ions present in the system. $L_z$ is the membrane thickness ($\sim$5 nm) and the coordinates of the ions were recorded at time intervals of $\Delta t=$1.5 ps. Water flux was induced by introducing a small force (0.01 kcal mol$^{-1}$nm$^{-1}$) to the oxygen atom of the water molecule. Other simulation parameters and protocols are discussed in Section 2.2.

The entry of the NP dendrimer into the pore lumen was rapid, whereas the entry of the P dendrimer was more gradual. After 200 ns, the P and NP dendrimers moved 15 {\AA} and 35 {\AA} into the pore lumen, respectively. Density maps revealed greater void space in case of the NP dendrimer~\cite{KGA-nanoscale}. Due to the strong electrostatic interactions between the negatively charged inner pore lumen and positively charged terminal groups of the P dendrimer, the P dendrimer blocked the pore lumen more effectively compared to the NP dendrimer, the latter adopting a stretched configuration (Fig.~\ref{Fig5}c). The ion distribution inside the pore was dramatically altered in the presence of the P dendrimer leading to significant reduction in ionic current with the P dendrimer (Fig.~\ref{Fig5}e and f). The total number of Cl$^{-}$ ions inside the nanopore also increased significantly due to increased binding of ions to the positively charged P dendrimer (Fig.~\ref{Fig5}b).

In the presence of the P dendrimer, residence time for Cl$^{-}$, evaluated by fitting the CSPs (Eq.~\ref{eq:CSP}) to a double exponential function, was 135 ps and 979 ps. The bound Cl$^{-}$ ions exhibited a long relaxation time of $\sim$ 1 ns, with the smaller relaxation time arising from the free ions within the pore. K$^{+}$ ions were absent in this region. This indicated strong attraction of Cl$^{-}$ and repulsion of K$^{+}$ ions by the P dendrimer. On the other hand, the NP dendrimer did not show any significant effect on the ions when compared to the free pore. This case study illustrates the value of MD simulations for selecting suitable pore blockers to potentially mitigate PFT-mediated infections.\newline

\section{MARTINI coarse-grained MD simulations of PFTs}

Many biological processes associated with PFTs such as oligomerization, membrane-reorganization, and lipid dynamics, can only be explored \textit{in silico} with large MD simulations accessing micro-meter and micro-second spatio-temporal scales (Fig.~\ref{Fig1}). A popular biomolecular CG FF is the MARTINI model{~\cite{martini-review,Martini_comp_microscopy,Bruininks2019,Pitfalls2019,martini,MartiniFF}}, whose chemical building blocks are primarily parameterized to reproduce experimental thermodynamic data such as oil/water partition coefficients. MARTINI is versatile, modular, and contains well-tested parameter sets for biomolecules such as proteins{~\cite{martini,MartiniFF,martini-protein,martini-improved_protein,elnedyn}}, lipids{~\cite{MartiniLipids0,Martini_improv_chol}}, and nucleic acids{~\cite{Martini_DNA,Martini_RNA}}.

\subsection{System preparation and simulation settings}

MARTINI simulations of PFTs~\cite{Desikan2017,Desikan_pnas_2020} were performed using FF version 2.2 with an elastic dynamic network (Elnedyn 2.2) implemented in GROMACS{~\cite{martini,MartiniFF,martini-protein,martini-improved_protein,Martini-PW,elnedyn}}. Firstly, the atomistic PFT protein structures were converted into CG resolution – 4 heavy atoms : 1 CG bead with standard amino acid mapping – using the `martinize.py’ tool; version 2.6 implemented in python 3, available from the official MARTINI website at \url{http://cgmartini.nl/index.php/tools2/proteins-and-bilayers} (accessed on 16\textsuperscript{th} October 2020). Since MARTINI simulations do not allow for secondary structure transformations, the initial secondary structure of the protein backbone, computed using the DSSP algorithm{~\cite{dssp}} (version 2.1.0), must be passed as an input into martinize.py for defining secondary structure constraints to the protein topology. We also employed default Elnedyn force constants{~\cite{elnedyn}} of 500 kJ mol\textsuperscript{-1} nm\textsuperscript{-2}, and note that while modifying this parameter to match global structural properties from atomistic simulations is possible{~\cite{deserno-elnedyn}}, we did not pursue that aspect in our work. With PFT oligomers and pores, an important detail to note is that the elastic network was applied within each protomer sub-unit in the oligomer/pore, and not across protomers. This is because the oligomeric assembly is supposed to be stabilized by non-bonded interactions, and applying an elastic network of springs across protomers could artificially stiffen/stabilize the oligomeric complexes.

The CG resolution transformed proteins were then inserted into homogeneous MARTINI DMPC, POPC, and DPPC (1,2-dipalmitoyl-sn-glycero-3-phosphocholine) bilayers, which were either created by self-assembly MD simulations, or by using the ‘INSANE’ python script{~\cite{Martini_insane}} (available from the same link as above), followed by short equilibration. During insertion of PFTs into membranes, the lipids overlapping with the protein were deleted similar to atomistic simulations (Section 2.3.3). Sufficient hydration was maintained by adding appropriate amount of CG water along with 0 or 0.15 M Na$^{+}$/Cl$^{-}$ ions to create charge neutral systems. Anti-freeze solvent particles, which have bigger van der Waals radii to disrupt artificial freezing of CG water, were sometimes added (proportion: 10\% of CG water beads). We note that contrary to typical observations, freezing was not observed with CG water in our PFT simulations even in the absence of anti-freeze beads, probably due to the heterogeneous nature of our membrane-protein-solvent systems (a large part of the PFTs are completely solvated), and the large numbers of CG ions disrupting freezing of the solvent.
	
The MARTINI simulation settings were fixed according to published parameter settings for improved performance in GROMACS{~\cite{Martini_mdp}} (sample input scripts available at \url{http://cgmartini.nl/index.php/force-field-parameters/input-parameters}, accessed on 16\textsuperscript{th} October 2020). While the MARTINI developers advocate an integration timestep of up to 40 fs, we conservatively ran all our MARTINI simulations with a 20 fs timestep for better accuracy and energy conservation. Simulations were performed in the NPT ensemble with weak temperature and pressure coupling enabled by the stochastic velocity-rescaling thermostat{~\cite{vrescale}} at 310 K (coupling constant of 1.0 ps), and the Parrinello-Rahman isotropic/semi-isotropic barostat{~\cite{parrinello}} at 1 bar (coupling constant of 12.0 ps). Electrostatic interactions were calculated using the reaction-field algorithm with a cut-off of 1.1 nm, and with a relative dielectric constant of 2.5 and 15 for systems with and without MARTINI polarizable water{~\cite{Martini-PW}}, respectively. The relative dielectric constant of the reaction field was set to infinity. van der Waal interactions were computed with a cut-off of 1.1 nm, and the Lennard-Jones potentials were shifted such that their values were zero at the cut-off. Upon energy minimization with steepest descent, short equilibration MD simulations with and without constraints on the membrane-protein beads were performed similar to atomistic simulations (see Section 2.2) before embarking on longer production simulations.

\subsection{Case studies}

\subsubsection{Protomer-membrane binding energy via umbrella sampling}

\begin{figure}[!ht]
\centerline{\includegraphics[scale=1]{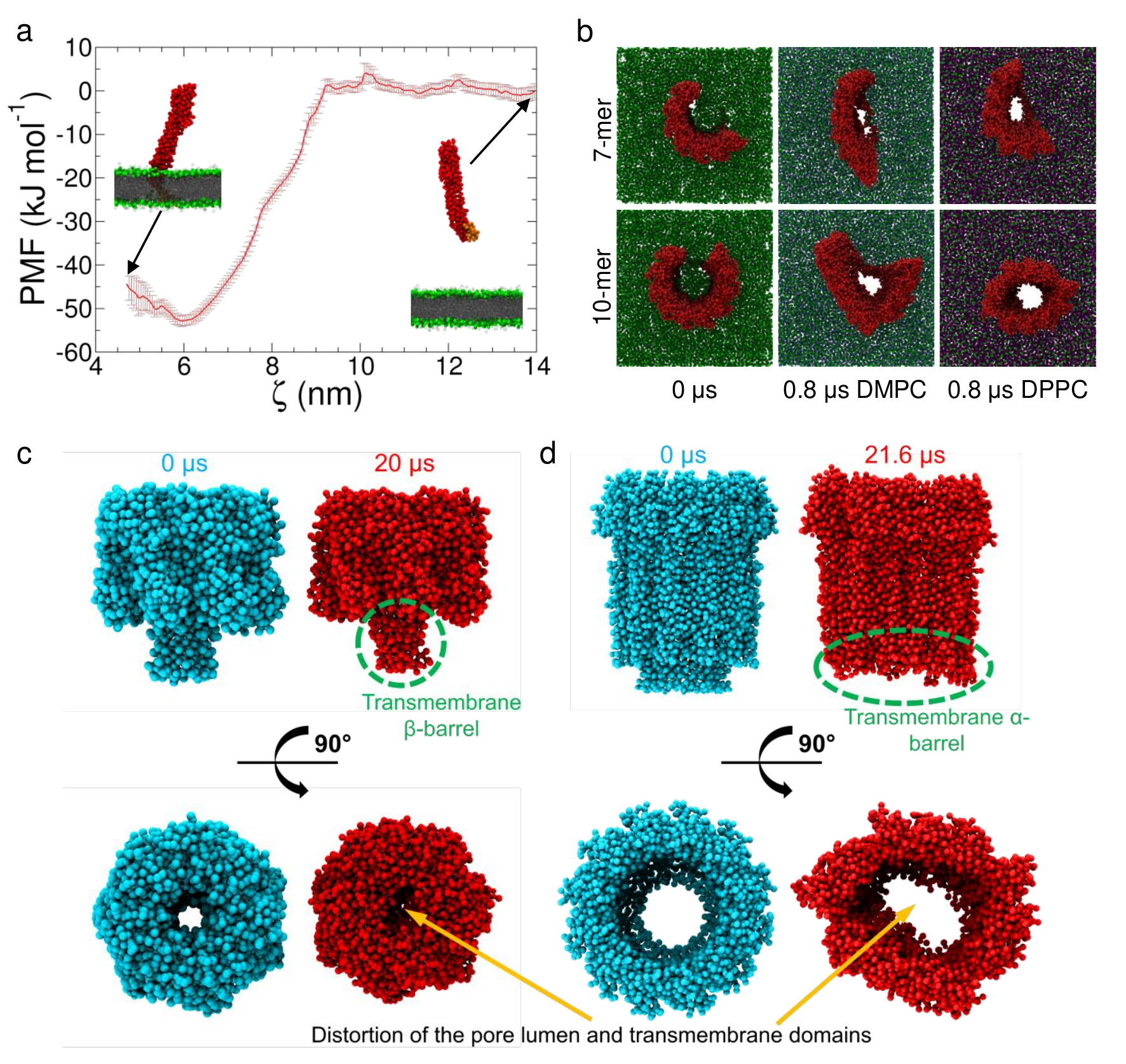}}
\caption{Martini model simulations for (\textbf{a}) potential of mean force (PMF) computations of a ClyA protomer with a phospholipid (DMPC) membrane~\cite{rajat-langmuir}, (\textbf{b}) lipid evacuation in membrane-inserted ClyA arcs~\cite{Desikan2017} (similar to Fig.~\ref{Fig3}b), and distortions in membrane-inserted pore complexes of (\textbf{c}) AHL and (\textbf{d}) ClyA pores~\cite{Desikan2017}. Only protein atoms shown for clarity in (c,d). (\textbf{a}) Reprinted with permission
from {\it Langmuir} {\bf 33}, 11496-11510, (2017). Copyright (2017) American Chemical Society. (\textbf{b-d}) Reprinted with Permission from {\it J. Chem. Sci.}, {\bf 129}, 1017-1030 (2017), Copyright (2017) Springer Nature.}
\label{Fig6}
\end{figure}

We employed MARTINI simulations, with a polarizable water model and PME electrostatics, to assess the binding free energy of a single ClyA protomer with a DMPC membrane using umbrella sampling simulations (Fig.~\ref{Fig6})~\cite{rajat-langmuir}. The reaction coordinate, `$\zeta$’, was defined as the distance between the centers of mass of the protomer and the membrane along the membrane normal. The ClyA protomer was equilibrated in the DMPC membrane for 100 ns, and subsequently, short steered MD simulations were utilized to create 24 starting configurations for umbrella sampling along $\zeta$, with $\Delta \zeta = 0.2$ nm. Each of these configurations were restrained around their respective initial values of $\zeta$ with weak harmonic potentials. From the variance of $\zeta$ in independent unrestrained simulations ($\zeta^*$), the spring constant for these weak harmonic restraints was calculated as $\frac{RT}{\zeta^*} = 10$ kJ mol\textsuperscript{-1} nm\textsuperscript{-2} at 310 K (see Ref.~\cite{rajat-langmuir} for more details). Movement of the membrane along its normal was restrained by applying a strong harmonic potential on the DMPC phosphate atoms (force constant of 100000 kJ mol\textsuperscript{-1} nm\textsuperscript{-2}). Each umbrella sampling window was then simulated for 100 ns. From examining the composite histogram plot along $\zeta$ from all umbrella sampling trajectories, it was observed that umbrella sampling windows 11 and 12, corresponding to the protomer at the water--membrane interface, exhibited large fluctuations and poor sampling (overlap). Thus, only these windows were rerun with a higher force constant of 100 kJ mol\textsuperscript{-1} nm\textsuperscript{-2}. Note that the N-terminus of the protomer, which is a part of ClyA's transmembrane domain, undergoes significant conformational changes in the fully solvated state (umbrella sampling simulations at higher $\zeta$). The potential of mean force (PMF) was obtained by using the weighted histogram analysis method (WHAM) along $\zeta$, and error bars on the PMF were computed by bootstrapping. From the PMF (Fig.~\ref{Fig5}a), the membrane-inserted protomer state is observed to be a thermodynamically favourable state, with a free energy difference of $\sim$ -13 kcal/mol compared to the fully solvated protomer.

\subsubsection{MARTINI simulations of PFT pores exhibit significant structural distortions}

PFT oligomers and pore complexes are stabilized by non-bonded interactions{~\cite{Desikan_jbsd_2020,mueller}}. While MARTINI FFs constrain the protein secondary structure, domain movements leading to changes in tertiary and quaternary structure are unrestricted. Hence, MARTINI simulations can be employed to assess the structural stability of PFT oligomers over longer timescales of tens of $\mu$s.

Lipid evacuation and stability of arcs were studied in MARTINI simulations. Indeed, while n-mer arcs evacuated lipids from their interior into the surrounding membrane, in agreement with atomistic simulations (Fig.~\ref{Fig6}b), they exhibited structural distortions in their transmembrane domains and exhibited a greater tendency to close with increasing number of protomers in the arcs (seen in the 10-mer arc in DPPC, Fig.~\ref{Fig6}b; more details in Ref.{~\cite{Desikan2017}}). In contrast, ClyA arcs in all-atom simulations show no tendency to close{~\cite{rajat-langmuir}}, and there is sufficient experimental evidence that stable ClyA arcs may be natural intermediates along ClyA (and $\alpha$-PFT) pore formation pathways~\cite{Ayush2017,peng2019high}. 

We also analysed the stability of ClyA and AHL membrane-inserted pores in MARTINI simulations~\cite{Desikan2017}. We observed that the quaternary structure of the solvated extra-cellular regions was well-preserved, but the transmembrane domains of both PFT pores were severely distorted (Fig.~\ref{Fig6}c and d), despite the application of an intra-protomer elastic network. Particularly, the transmembrane domain of the AHL pore was constricted to such an extent that the pore channel was nearly closed, which would essentially render the pore non-functional. This is in stark contrast to experimental observations, crystal structures, and all-atom simulations where the pore geometries are stable and in close agreement with crystal structures (albeit performed over shorter time scales of $\sim$1000 ns; Refs.{~\cite{Desikan_jbsd_2020,Desikan2017}}).

There are a few potential remedies to stabilize the tertiary and quaternary structures of protein complexes in MARTINI simulations. It has been shown that additional inter-protomer elastic dynamic networks may stabilize the entire arc/pore complex{~\cite{Vogele2019,Basdevant2019}}. Also, force constants of the elastic network can be calibrated using all-atom simulations~\cite{deserno-elnedyn}. Finally, we note that the next generation MARTINI 3 FFs currently in development (see \url{http://cgmartini.nl/index.php/martini3beta}, accessed on 16\textsuperscript{th} October 2020), with new particle types, reparametrized interaction matrix, and improved representation of proteins, may improve the stability of pore complexes in MARTINI simulations.\newline

\section{Structure-based models for capturing the membrane-triggered conformational transition of an $\alpha$-PFT monomer}

A unique feature of PFTs is that their sequence encodes multiple 3D conformations that are stable depending on the environment{~\cite{Goot-PFTreview2015}}. In the case of ClyA, the secreted water-soluble monomeric form undergoes a large conformational change upon membrane recognition and binding into its assembly-competent protomeric state{~\cite{mueller,rajat-smog}}. This conformational transition involves $\beta$-strand to $\alpha$-helix and loop to $\alpha$-helix secondary structure transformations, reorganization of inter-helix interfaces, and domain rearrangements of up to 140 {\AA}, making it one of the largest conformational transitions observed in proteins of length of about 300 amino acids~\cite{rajat-smog,mueller}. The timescale of this process is estimated to lie in the range of seconds to minutes{~\cite{schuler-natcomm,vaidya}}, rendering it completely inaccessible to conventional atomistic molecular simulations, perhaps even with enhanced sampling techniques. Therefore, we employed MD simulations with a dual-basin, coarse-grained, implicit-solvent, structure-based model (SBM) of ClyA, along with a coarse-grained membrane model, to glean molecular insights into this conformational transition.

\begin{figure}[!ht]
\centerline{\includegraphics[scale=1]{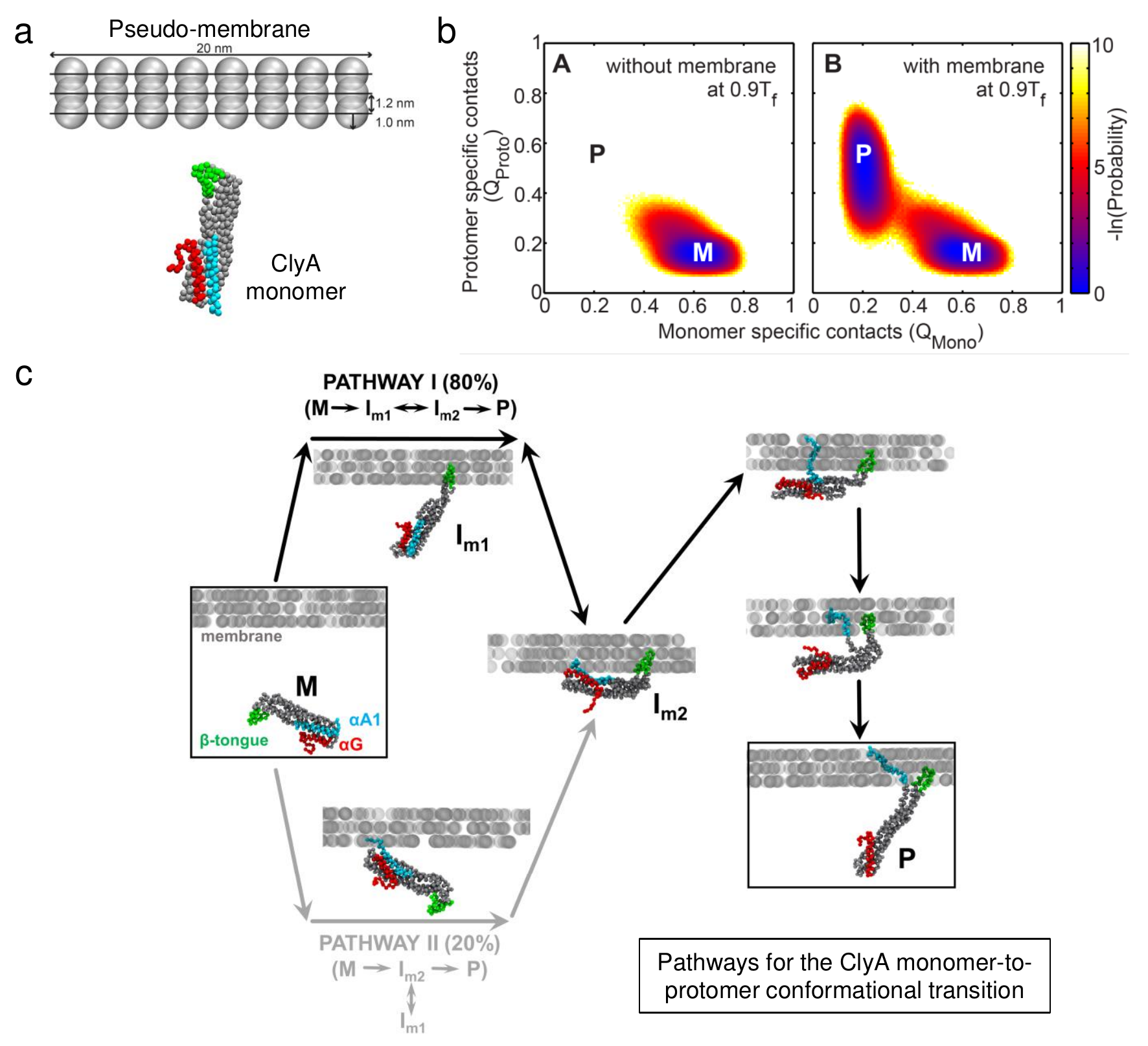}}
 \caption{Structure based models (SBMs) illustrate the membrane-triggered conformational transition of a ClyA monomer into a protomer during membrane-insertion\cite{rajat-smog}. (\textbf{a}) Initial SBM MD configuration with the ClyA monomer situated away from the membrane. (\textbf{b}) Transition to protomer observed only in presence of a membrane. (\textbf{c}) Pathways observed during membrane-insertion. Reprinted with permission
from {\it The Journal of Physical Chemistry B} {\bf 120}, 12064-12078, (2019). Copyright (2019) American Chemical Society.}
\label{Fig7}
\end{figure}

SBMs encode the native protein structure into a coarse-grained potential energy function (C$_\alpha$ protein atoms only) consisting of bonded (covalent bonds, angles, dihedrals) and non-bonded (modified contact potentials; no electrostatics) terms, which have been shown to capture the funnel-shaped energy landscape of proteins{~\cite{Clementi2000,SBM-bookchapter,Hemanth_pnas}}. To investigate the membrane-triggered PFT conformational transition, we employed dual-basin SBMs{~\cite{Okazaki2006,Whitford2007,Lu2008,RFAH-dSBM,LinPNAS}} (dSBMs), where both the ClyA monomer and protomer structures were encoded into the same potential energy function. Additionally, we also constructed a minimalist coarse-grained membrane model (Fig.~\ref{Fig7}a) that encoded essential membrane--ClyA protomer interactions compatible with C$_\alpha$ dSBMs. While the monomer conformation was expected to be stable in the absence of the membrane, interaction with the membrane ``triggered'' the transition into the protomer conformation (Fig.~\ref{Fig7}b) while conferring stabilizing interactions with the transmembrane domain~\cite{rajat-smog}. 

Indeed, dSBM simulations, starting with the monomer state away from the membrane (Fig.~\ref{Fig7}a), captured ClyA's membrane-triggered monomer-to-protomer conformational transition (Fig.~\ref{Fig7}b) via two pathways (Fig.~\ref{Fig7}c). A dominant serial pathway via two membrane-bound intermediates ($I_{m1}$ and $I_{m2}$) occurs in $\sim$80\% of the trajectories, while a secondary pathway where the intermediate $I_{m2}$ appears as an off-pathway product occurs in the remaining $\sim$20\% of the trajectories (Fig.~\ref{Fig7}c). In our simulations, this transition only occurred in the presence of a membrane{~\cite{rajat-cter}} (Fig.~\ref{Fig7}b). Our simulations showed good agreement with FRET experiments{~\cite{schuler-natcomm,ban-biochem}}, which served as validation. A crucial insight was that one of ClyA's membrane-interacting domains, the $\beta$-tongue, functioned as a quick-response membrane sensor with a small barrier for conformational change and membrane-insertion, while the N-terminus helix $\alpha$A1, functions as a fidelity enhancer to complete the conformational transition in the target membrane.\newline

\section{Summary}

Molecular dynamics simulations have evolved as a powerful \textit{in silico} framework to study a variety of biologically relevant molecules. In this article, we illustrate the utility of MD simulations to provide insights into the pore forming pathways of PFTs belonging to distinct PFT families, and their ability to provide molecular insights into biological phenomenon often difficult to probe experimentally. We have laid emphasis on methods and techniques involved while setting up, monitoring, and evaluating properties from MD simulations of PFTs in a membrane environment. Choosing the appropriate MD technique is governed by the phenomenon we seek to explore, as well as the ability to reliably sample ensemble interactions within typical sampling times associated with a given technique. Thus, all-atom MD simulations with appropriate force fields from the CHARMM and AMBER families are expected to accurately capture protein-protein, protein-lipid and protein-water interactions. Additionally, current computing resources provide the ability to reliably monitor phenomenon that occur on the order of microseconds for moderately sized pore complexes such as those associated with ClyA and AHL. Understanding conformational changes that occur over time scales of milliseconds to minutes requires the use of advanced sampling techniques or reduced models such as the SBMs. MARTINI models are another class of coarse grained methods have been used to study oligomers formed by PFTs. 

With rapid advances in super resolution and single molecule microscopy~\cite{sarangi2018,sarangi2016super,kga-pnas}, MD simulations are expected to play a significant role in interpreting experimental data and providing molecular insights into the PFT pathways as illustrated in this article. With the latest stimulated emission depletion (STED) microscopes, which can probe phenomenon down to 30 nm length scales accessible in MD simulations, we expect a strong synergy to occur between experiments and simulations. Similarly, with advances in Cryo-EM techniques~\cite{groll2018structurea,peng2019high}, high resolution crystal structures of PFT pore complexes as well as oligomeric intermediates in native membrane environments may be forthcoming, and these techniques may open up new frontiers by providing valuable inputs and insights to MD simulations of PFTs, and vice versa.\newline \newline

\section*{Acknowledgements}
This work was supported by the Department of Science and Technology, Science and Engineering Research Board. We thank the Supercomputer Education and Research Center, Indian Institute of Science for computational facilities.\newline \newline

\section*{Competing interests}

The authors declare no competing interests. \newline \newline

\bibliography{Refs}
\bibliographystyle{naturemag}

\end{document}